\documentclass{PoS}
\usepackage{graphicx}
\usepackage{amsmath}
\usepackage{aasmacros}
\usepackage{xspace}
\usepackage{url}

\newcommand{\hess}{H.E.S.S.\xspace}
\newcommand{\lat}{\textit{Fermi}-LAT\xspace}

\newcommand{\pks}{PKS\,2023$-$07\xspace}
\newcommand{\degr}{\ensuremath{^\circ}\xspace}

\title{Observations of the blazar \pks in flaring state with \hess and \lat in 2016--2017 and constraints on an intrinsic cut-off}

\ShortTitle{\pks in flaring state with \hess and \lat in 2016--2017}

\author{\speaker{G.~Emery}$^1$, F.~Jankowsky$^2$, J.-P.~Lenain$^1$, J.~P.~Marais$^3$, T.~Mbonani$^3$, C.~Romoli$^4$, B.~van~Soelen$^3$, A.~Wierzcholska$^{2,5}$, M.~Zacharias$^{6,7}$, for the \hess collaboration\footnote{for collaboration list see PoS(ICRC2019)1177}, M.~Meyer$^8$\\
        $^1$Sorbonne Universit\'e, Universit\'e Paris Diderot, Sorbonne Paris Cit\'e, CNRS/IN2P3, Laboratoire de Physique Nucl\'eaire et de Hautes Energies, LPNHE, 4 Place Jussieu, F-75252 Paris, France\\
        $^2$Landessternwarte, Universit\"at Heidelberg, K\"onigstuhl, D69117 Heidelberg, Germany\\
        $^3$Department of Physics, University of the Free State, 9300 Bloemfontein, South Africa\\
        $^4$Max Planck Institute f\"ur Kernphysik, Saupercheckweg 1, 69117, Heidelberg, Germany\\
        $^5$Instytut Fizyki Jadrowej PAN, ul. Radzikowskiego 152, 31-342 Krak\'ow, Poland\\
        $^6$Theoretische Physik IV, Ruhr-University Bochum, Germany\\
        $^7$Centre for Space Research, North-West University, Potchefstroom 2520, South Africa\\
        $^8$W. W. Hansen Experimental Physics Laboratory, KIPAC , Department of Physics and SLAC National Accelerator Laboratory, Stanford University, Stanford, CA 94305, USA\\
        E-mail: \email{gemery@lpnhe.in2p3.fr}$^*$}

\abstract{\pks is a flat spectrum radio quasar located at a redshift $z=1.388$, farther than any source currently detected at very high energies ($E>100$\,GeV). At such energies, absorption by the extragalactic background light (EBL) renders the detection of distant sources particularly challenging. The High Energy Stereoscopic System (\hess) observed the source following reports from AGILE (April 2016) and \lat (April 2016, September and October 2017) on high-flux states in gamma rays. During each of the three flaring periods, near-simultaneous observations were obtained with \hess, \lat and multiple telescopes at other wavelengths. Though the source was not significantly detected by \hess, upper limits were derived for each observation period. Through constraints given by \lat in the MeV--GeV domain and differential upper limits by \hess, we searched for an intrinsic cutoff in the EBL-corrected gamma ray spectrum of \pks. 
}

\FullConference{36th International Cosmic Ray Conference -ICRC2019-\\
                July 24th - August 1st, 2019 \\
                Madison, Wisconsin, USA}

\begin{document}
\section{Introduction}
Observations of extragalactic sources at very high energies (VHE, $E>100$\,GeV) is strongly limited by the distance of the sources in particular due to the absorption of VHE photons by the extragalactic background light (EBL), the light produced by galaxies, stars and dust during the Universe evolution~\cite{2001ARA&A..39..249H}. In this context, active galactic nuclei (AGN) were only detected at VHE with redshifts below 1~\cite{2017ApJ...837..144P}~\cite{2016A&A...595A..98A}. AGN are massive black holes located at the center of some galaxies which emit strongly over most of the electromagnetic spectrum while accreting matter. Around 10\% of AGN power a relativistic jet. If the jet points in Earth direction, the AGN is then called a blazar and VHE photons produced by the jet can be observed.

\pks is a flat spectrum radio quasar (FRSQ) (sub-category of blazars~\cite{2017A&ARv..25....2P}) located at a redshift of $z=1.388$. Even if \pks intrinsic emission at VHE were significant, absorption by EBL photons makes observations impossible in an average flux state. Since AGN are strongly variable, it should be possible to detect the source during an intense enough flaring activity.

The AGN Target of Opportunity (ToO) program of \hess searches for sources in high emission state seen by other experiments at all energies (optical, X-rays, HE, VHE) and triggers observations. For AGN otherwise too faint, it can allow for detection in a reasonable amount of time. And for stronger, potentially already detected AGN, time resolved and high quality data can be obtained.

\section{High and Very High Energy Observations}
\subsection{\hess observations}

\hess, the High Energy Stereoscopic System, is an array of telescopes located in Namibia. The array is composed of four telescopes with a mirror diameter of 12 meters called CT1 to CT4, and one with a mirror diameter of 28 meters called CT5.
When a VHE photon reaches the atmosphere it interacts creating an electromagnetic shower with enough energy to produce Cherenkov light. The elliptical images obtained on the camera can be combined using the stereoscopic technique to obtain the energy and arrival direction of the VHE photon in a 5 degrees diameter field of view. \hess can also operate in monoscopic mode using using only events detected with the CT5 telescope. In monoscopic mode, the field of view is reduced to 2.5 degrees of diameter.

\hess observations were triggered following a high state detected with AGILE~\cite{2016ATel.8879....1P} and \lat~\cite{2016ATel.8932....1C} in April 2016, and in September 2017 and October 2017 following private communication from the \lat team and alerts issued with \texttt{FLaapLUC}~\cite{2018A&C....22....9L}.

The April 2016 flare was observed by \hess for 56 minutes at the start of the flare. The observations were limited by the short observation window available from the \hess site at the time and technical issues.
During the September 2017 flare, 155 minutes of observation were possible at the start of the flare and 358 minutes at the end. The peak of the high energy flare couldn't be observed by \hess due to technical issues.
In October 2017, the decreasing phase of the high energy flare was observed for 336 minutes.
Detailed observation times are displayed on Fig.\ref{fig1}-a).

Analysis of the data taken for each flare was performed in mono mode using CT5 only. CT5 with its larger collection area possess a lower energy threshold than CT1-4 and is hence the most likely to detect \pks since absorption by the EBL is more efficient at higher energies. Each analysis was performed with two independent calibration and analysis chains to ensure robustness of the results, here the Model reconstruction~\cite{2009APh....32..231D} was used. Compatible results were obtained using the image template fitting method (ImPACT)~\cite{2015arXiv150900794M}.
No detection was possible for each periods and only upper limits on the flux could be derived. The $95\%$ C.L. differential upper limits obtained assuming a spectral index of -3 are shown on Fig.\ref{fig2}.
The $95\%$ C.L. integral upper limits above threshold are also produced with the same spectral hypothesis. The difference in energy threshold is due to different zenith of observation. For each flare the zenith, energy threshold and integral upper limit are compiled in Table~\ref{table1}.

\begin{table}
\resizebox{0.8\textwidth}{!}{%
\hspace{0.2\textwidth}\begin{tabular}{|c|c|c|c|}
\hline
Flare & Zenith & $E_{threshold}$ & Flux(> $E_{threshold})$ ($ph.cm^{-2}.s^{-1}$)\\
\hline
April 2016  & 37.0 & 108 GeV & < 3.3 $\times$ $10^{-11}$ \\
\hline
September 2017 & 20.1 & 73 GeV & < 5.2 $\times$ $10^{-12}$ \\
\hline
October 2017  & 24.6 & 73 GeV & < 9.1 $\times$ $10^{-12}$ \\
\hline
\end{tabular}
}
\caption{\hess zenith of observation, energy threshold and integral upper limits}
\label{table1}
\end{table}

\subsection{\textit{Fermi}-LAT observations}

\textit{Fermi} is a space-based gamma-ray observatory that orbits the Earth since June 2008. Its main instrument, the Large Area Telescope (LAT) sees the whole sky every two orbits (i.e. every $\sim$3\,h), thanks to a large field of view (2.4\,sr). \lat is sensitive to HE photons from $\sim$20\,MeV to $>$300\,GeV. For most of its operation strategy, it operates in survey mode, which guarantees a complete sky survey in a sub-day time-range, while providing a fair exposure for relatively bright sources~\cite{2009ApJ...697.1071A}.

\lat data were analysed in several subsets, in order to have contemporaneous results with the HESS observations (Table \ref{table2}) for the three flares. We used the standard \textit{FermiTools}\footnote{\url{https://github.com/fermi-lat/Fermitools-conda}} software version \texttt{1.0.0} and the Pass~8 event selection (event class and instrument response functions \texttt{P8R3\_SOURCE\_V2}). For each flare, spectra and light curves were derived using the binned likelihood analysis \texttt{gtlike} package in the energy range 100\,MeV to 500\,GeV. The region of interest is 10\degr of radius, and the recommended selection of time intervals are used (\texttt{DATA\_QUAL>0 \&\& LAT\_CONFIG=1}). Events are selected with a zenith angle below 90\degr. The source input model was built based on the 3FGL catalogue, using the \texttt{make3FGLxml} user-contributed script, and includes the Galactic interstellar emission model (\texttt{gll\_iem\_v06}) and the relative isotropic diffuse emission template (\texttt{iso\_P8R3\_SOURCE\_V2}). The likelihood fit is performed iteratively: first, the model is used as per the 3FGL catalogue. In a second step, sources for which TS$<$9 and contributing to less than 5\% of the event counts in the whole data set have all their parameters fixed. Finally, in a third iteration, the parameters of all sources beyond 3\degr from \pks are fixed.
Considering the short time intervals for the flares presented in this study, we assume a power-law shape for the spectrum of \pks, even though \pks is best described with a log-parabola spectrum in the 3FGL catalogue~\cite{2015ApJS..218...23A}\footnote{As well as in the 4FGL catalogue~\cite{2019arXiv190210045T}.}. 
We checked with a log-likelihood ratio test for each flare period that a log-parabolic spectral shape does not improve significantly the fit results. The most intense flare was found to happen in April 2016 but was not the most covered with \hess The daily binned light curve is displayed on Fig.\ref{fig1}-a).
\begin{table}
\resizebox{\textwidth}{!}{%
\begin{tabular}{|c|c|c|c|c|c|c|c|}
\hline
Flare  & Calendar date   & MJD & MET & duration & Model & Flux 0.1$-$500 GeV  & Signif\\
       &                 &      &   & (h)       &          &        ($ph.cm^{-2}.s^{-1}$)     &  \\        
\hline
apr16  & 12 Apr 2016 at 12:00$-$14 Apr 2016 at 12:00 & 57490.50$-$57492.50 & 482155204$-$482328004 & 48.0 & PL & (2.53 $\pm$ 0.21) $\times$ $10^{-6}$ & 30.7 \\
\hline
sept17-a & 12 Sep 2017 at 07:00$-$13 Sep 2017 at 12:46 & 58008.29$-$58009.53 & 526892408$-$526999588 & & & & \\
sept17-b & 19 Sep 2017 at 06:33$-$20 Sep 2017 at 12:47 & 58015.27$-$58016.53 & 527495609$-$527604445 & 108.8 & PL & (5.77 $\pm$ 0.87) $\times$ $10^{-7}$ & 17.0 \\
sept17-c & 21 Sep 2017 at 06:34$-$23 Sep 2017 at 07:24 & 58017.27$-$58019.31 & 527668446$-$527844258 & & & & \\
\hline
oct17  & 12 Oct 2017 at 12:00$-$15 Oct 2017 at 12:00 & 58038.50$-$58041.50 & 529502405$-$529761605 & 72.0 & PL & (3.12 $\pm$ 0.79) $\times$ $10^{-7}$ & 7.9 \\
\hline
\end{tabular}
}
\caption{\lat spectral analysis}
\label{table2}
\end{table}

\section{Multi-Wavelength campaign}

AGILE is a space mission containing two instruments: a gamma-ray detector, sensitive to photons with energy in the range 30\,MeV--50\,GeV, and a hard X-ray detector, sensitive in the range 18--60\,keV. AGILE reported a flux of $(2.3 \pm 0.8) \times 10^{-6}$\,ph cm$^{-2}$\,s$^{-1}$ ($E > 100$\,MeV) integrated from 2016-03-25 12:00 UT to 2016-03-27 12:00 UT.

\begin{figure*}
  \includegraphics[width=1\textwidth]{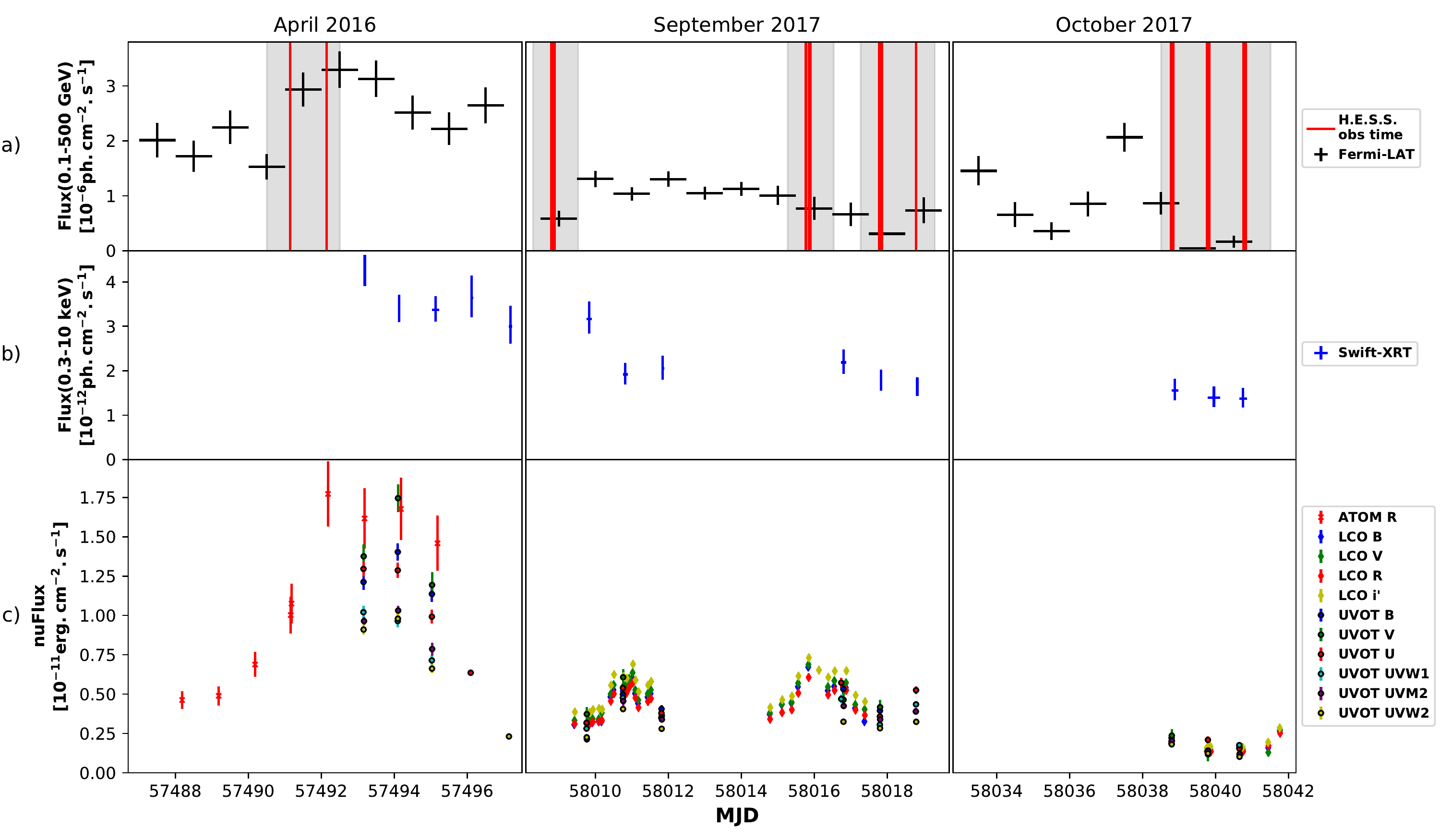}
  \caption{Multi-wavelength observations of PKS 2023-07.
    \textbf{a)} \lat daily binned light curve integrated between 100 MeV and 500 GeV. Red lines represent \hess observations and the shaded grey areas cover the time windows used for the \lat spectral analysis (shown in Fig.\ref{fig2}).
    \textbf{b)} X-rays light curve with \textit{Swift}-XRT integrated between 0.3 and 10 keV.
    \textbf{c)} optical fluxes with ATOM and LCO and optical/UV fluxes with \textit{Swift}-UVOT.}
\label{fig1}
\end{figure*}

The Neil Gehrels \textit{Swift} observatory~\cite{2004ApJ...611.1005G} is comprised of three instruments, among which the XRT~\cite{2005SSRv..120..165B} detects X-rays between 0.2 and 10\,keV, and the UVOT observes at UV and optical wavelengths between 170 and 600\,nm. Observations were triggered following the \hess observations.
XRT data, which were all acquired in the standard photon counting mode, are analysed using the \texttt{HEASoft} suite version 6.22.1. Events are cleaned using the standard criteria from \texttt{xrtpipeline}. Data from \pks are extracted within a circle of 20 pixels, and the background is taken from an annular region with an inner radius of 50 pixels and outer radius of 160 pixels.
\texttt{XSpec} version 12.9.1p is used for the spectral analyses, where events below 0.3\,keV are excluded. Accounting for a Galactic hydrogen column of $3.24 \times 10^{20}$\,cm$^{-2}$~\cite{2005A&A...440..775K}, energy flux measurements, assuming a power-law spectral shape, in the energy range 0.3--10\,keV for each exposure contemporaneous with \hess observations are reported on Fig.\ref{fig1}-b).
Simultaneously with XRT instrument, PKS\,2023-07 was monitored in the ultraviolet and optical bands with the UVOT one.
All ultraviolet and optical magnitudes and fluxes have been calculated using \verb|uvotsource| procedure including all photons  from a circular region with radius 5''.
The background was determined as a circular region with a radius of 10''.
UVOT light curves are on Fig.\ref{fig1}-c). 
All data points are corrected for dust absorption.

The Automatic Telescope for Optical Monitoring (ATOM) is an optical telescope located at the H.E.S.S. site in Namibia. It provides optical monitoring on known gamma-ray emitters as well as multi-wavelength support for target-of-opportunity events and covered the 2016 flare event in R band. Data was reduced and analysed using ADRAS 2.4.14. Fluxes are obtained via differential photometry using between three and six custom calibrated comparison stars. The associated lightcurve is shown on Fig.\ref{fig1}-c).


The {\it Las Cumbres Observatory} (LCO) is a worldwide network of robotic telescopes. During the 2017 flares, LCO observed the target in B,V,R,i$^\prime$. Pre-reduction was performed with the LCO pipeline and differential photometry was performed using {\sc PyRAF}.  Differential photometry was performed using six comparison stars in the same field of view with magnitudes derived from the PanSTARR1 survey~\cite{2016arXiv161205560C}. The light curve is shown on Fig.\ref{fig1}-c).

\section{Discussion}

\begin{figure}
\hspace{-0.01\textwidth}
  \begin{minipage}[t]{0.5\textwidth}
    \includegraphics[width=\textwidth,keepaspectratio]{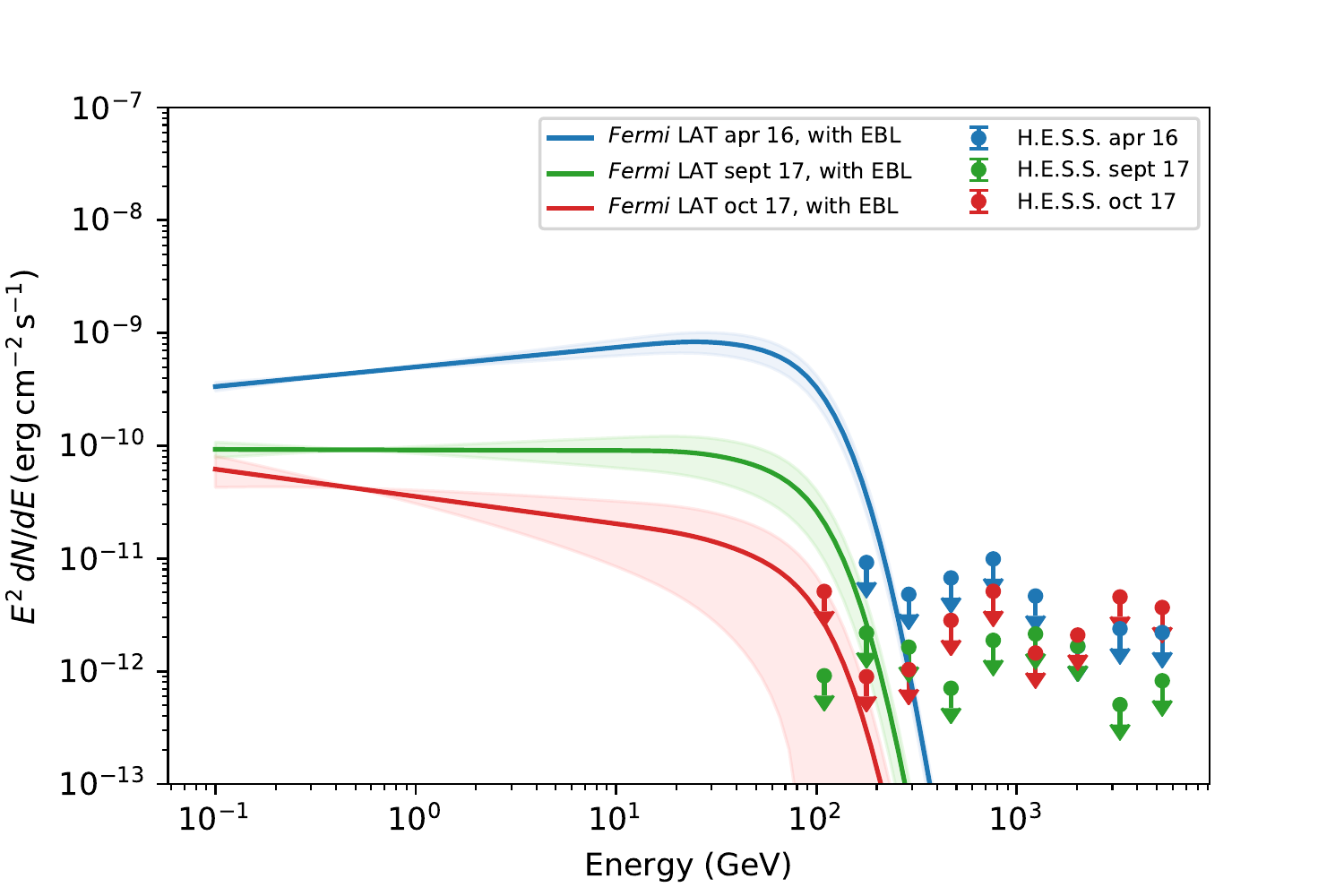}
    \caption{\hess $95\%$ C.L. differential upper limits obtained for each observation campaigns. Contemporaneous \lat power law spectra obtained from the greyed period on Fig.\ref{fig1}-a) (see also Table \ref{table2}) and extrapolated at VHE with the same power law corrected by the EBL absorption following the model by Dominguez et al.~\cite{2011MNRAS.410.2556D} are also displayed.}
    \label{fig2}
  \end{minipage}
\hspace{0.02\textwidth}
  \begin{minipage}[t]{0.5\textwidth}
    \includegraphics[width=\textwidth,keepaspectratio]{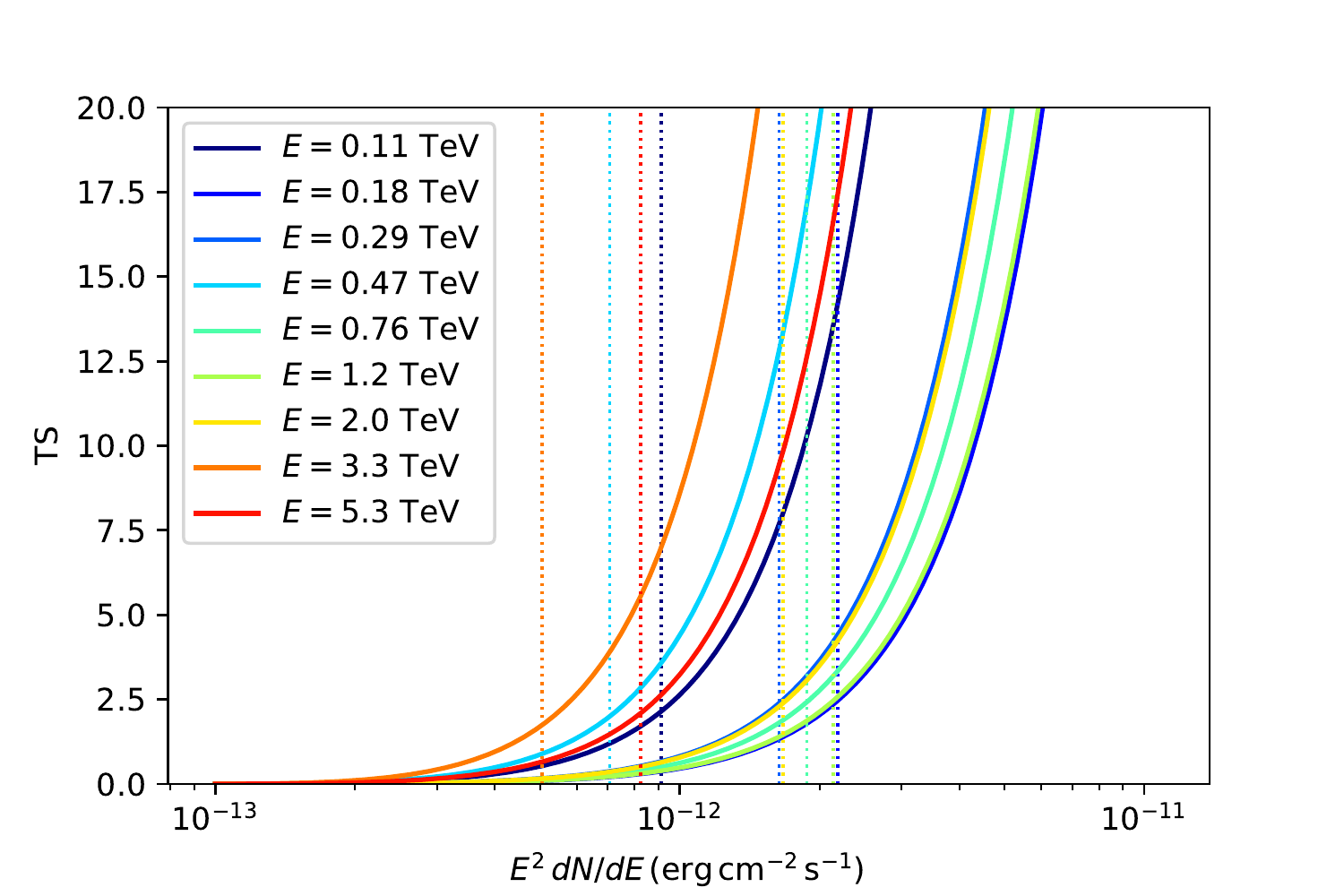}
    \caption{Test Statistic of the probability of a detection by H.E.S.S for each differential upper limit point of the September flare as a function of the flux of the source at the associated energy. The dot-lines correspond to the flux giving a detection with a probability of 95\% for the associated upper limit.}
    \label{fig3}
  \end{minipage}
\end{figure}

Extrapolating the \lat power-law spectra near-simultaneous with \hess observations for each flare (taking all \lat observations in the grey areas on Fig.\ref{fig1}-a), also detailed in Table~\ref{table2}) and taking into account absorption by the EBL with the model by Dominguez et al.~\cite{2011MNRAS.410.2556D}, we can compare them with the \hess differential upper limits (Fig.\ref{fig2}). The 95\% C.L. upper limits at VHE probe lower fluxes than the \lat extrapolated spectra.
Clearly, a simple power-law extrapolation and standard EBL absorption are not compatible with the non-detection in April and September, whereas the 
\hess upper limits for October can be explained with a simple power-law extrapolation of the intrinsic blazar spectrum. 

To explain the non-detection by \hess, three hypothesis are tested independently. The first is the potential presence of an intrinsic cut-off in the emission region. The second is a different normalisation of the EBL than the one from the model used in the extrapolations. The last is the absorption of the VHE photons in the broad line region (BLR) around the AGN.
All constraints are reported at one-sided 95\,\% confidence levels.

To do so, we assume that the \hess measurements are Gaussian random variables and accordingly, the \hess upper limits in each energy bin can be translated into a Gaussian likelihood profile, see Fig.\ref{fig3}. 
Then, for a given extrapolation of the \lat SED, the predicted flux level is used in each energy band to obtain the summed likelihood. Using a profile likelihood ratio test~\cite{2005NIMPA.551..493R}, we can then evaluate if a given extrapolation is compatible with the \hess data. 

First, we vary the normalization, $n_\mathrm{EBL}$, of the EBL photon density (modeled with model of Ref.\cite{2011MNRAS.410.2556D}) and compute the profile likelihood as a function of $n_\mathrm{EBL}$. We find that one would have to increase the EBL density by at least a factor of $3.07$ and $1.39$ for the September 2017 and April 2016 observation campaigns, respectively. 
Such high levels of the EBL are not compatible with luminosity functions of galaxy and other gamma-ray measurements~\cite{2011MNRAS.410.2556D,2018Sci...362.1031F}.

Next, we test the hypothesis that the intrinsic spectrum is a power law modified with an exponential cut-off at energy $E_\mathrm{cut}$.
Again computing the profile likelihood this time as a function of $E_\mathrm{cut}$ we find that $E_\mathrm{cut} < 128\ \text{GeV} $ for the April 2016 flare and $E_\mathrm{cut} < 35\ \text{GeV}$ for the September 2017 flare. 
In a scenario where very high-energy gamma rays are produced in inverse Compton scattering of electrons with external radiation fields, as commonly assumed for blazars~\cite{2016ARA&A..54..725M}, this would imply a cut-off or break in the electron distribution at a Lorentz factor in the comoving frame
$\gamma' = \delta_\mathrm{D}^{-1}\sqrt{E_\mathrm{cut}(1+z)/(2E_0)}$ if the electrons scatter predominantly off photons at energy $E_0$ in the Thompson regime~\cite{2016ApJ...830...94F} and where $\delta_\mathrm{D}$ is the Doppler factor.
Assuming $\delta_\mathrm{D} = 20$, we arrive at $\gamma' \lesssim 3.2\times 10^3 $ for scattering off BLR photons at 10\,eV, $\lesssim 1.4\times 10^3$ for scattering with accretion disk photons at 50\,eV, and $\lesssim 3.2\times10^4$ for scattering with photons of the dust torus at a temperature of 1000\,K for $E_\mathrm{cut} < 35\ \text{GeV}$.

Lastly, we test if the cut-off could be due to the interaction of VHE gamma-rays with BLR photons. Following Ref.\cite{2019arXiv190202291M}, we use the BLR model of Ref.\cite{2016ApJ...830...94F}, where we assume the ring geometry, and calculate the profile likelihood as a function of the distance $r$ of the gamma-ray emitting region to the central supermassive black hole. 
Searching the literature, we did not find estimates of the disk or H$\beta$ luminosity and neither for black hole mass of \pks. Therefore, we assume $L_\mathrm{disk} = 10^{46}\,\mathrm{ergs}.\mathrm{s}^{-1}$ and $L(\mathrm{H}\beta) = 10^{43}\,\mathrm{ergs}.\mathrm{s}^{-1}$, which corresponds to a total BLR luminosity of $0.03 L_\mathrm{disk}$ using the scaling relations of Ref.\cite{2016ApJ...830...94F}, and a black hole mass $M_\bullet = 10^9\,M_\odot$, where $M_\odot$ is the solar mass.
The \hess data are compatible with a power-law extrapolation of the LAT spectrum and BLR absorption if the emission region is closer than 
 $r = 1.8\times10^{17}\,\text{cm} \sim 120\,r_g$ ($r = 9.5\times10^{17}\,\text{cm} \sim 650\,r_g$) for the April 2016 (September 2017) flare, where $r_g = GM_\bullet/c^2$ is the gravitational radius with the gravitational constant $G$ and the speed of light $c$.

\section{Conclusion}

After the detection of bright activity phases at high-energy gamma rays, \hess observed \pks in three occasions in April 2016, and September and October 2017 leading to no detection at very high energies. 
Even with the strong EBL absorption expected at a redshift of 1.388, the extrapolation of the high-energy gamma-ray spectra is incompatible with the upper limits derived for the April 2016 and September 2017 flares. 
A correction to the EBL density compared to the model by Dominguez et al. is ruled out due to the large correction factor which would be needed and is incompatible with the measured values.
Rather, the non-detection is either caused by a break in the emitted spectrum or absorption of VHE photons in the BLR if the emission region is closer to the central black hole than several hundreds of gravitational radii.

\section{Acknowledgements}

\noindent The full H.E.S.S. acknowledgements can be found at the following link :\\ https://www.mpi-hd.mpg.de/hfm/HESS/pages/publications/auxiliary/\\HESS-Acknowledgements-2019.html\\


\noindent We acknowledge the use of public \lat data and the input from Dr. Richard J. Britto and Dr. Sara Buson. We warmly thank the Neil Gehrels \textit{Swift} Observatory team for the approval and prompt scheduling of ToO observations. We acknowledge the contribution of the ATOM collaboration and use of ATOM data. We acknowledge the contribution of the LCO collaboration and use of LCO data. A.W. is supported by Polish National Agency for Academic Exchange (NAWA).


\bibliographystyle{JHEP}
\bibliography{bibliography}

\providecommand{\href}[2]{#2}\begingroup\raggedright\begin{thebibliography}{10}

\bibitem{2001ARA&A..39..249H}
M.~G. {Hauser} and E.~{Dwek}, \emph{{The Cosmic Infrared Background:
  Measurements and Implications}},
  \href{https://doi.org/10.1146/annurev.astro.39.1.249}{\emph{\araa} {\bfseries
  39} (2001) 249} [\href{https://arxiv.org/abs/astro-ph/0105539}{{\ttfamily
  astro-ph/0105539}}].

\bibitem{2017ApJ...837..144P}
S.~{Paiano}, M.~{Landoni}, R.~{Falomo}, A.~{Treves}, R.~{Scarpa} and
  C.~{Righi}, \emph{{On the Redshift of TeV BL Lac Objects}},
  \href{https://doi.org/10.3847/1538-4357/837/2/144}{\emph{\apj} {\bfseries
  837} (2017) 144} [\href{https://arxiv.org/abs/1701.04305}{{\ttfamily
  1701.04305}}].

\bibitem{2016A&A...595A..98A}
M.~L. {Ahnen}, S.~{Ansoldi}, L.~A. {Antonelli}, P.~{Antoranz}, C.~{Arcaro},
  A.~{Babic} et~al., \emph{{Detection of very high energy gamma-ray emission
  from the gravitationally lensed blazar QSO B0218+357 with the MAGIC
  telescopes}}, \href{https://doi.org/10.1051/0004-6361/201629461}{\emph{\aap}
  {\bfseries 595} (2016) A98}
  [\href{https://arxiv.org/abs/1609.01095}{{\ttfamily 1609.01095}}].

\bibitem{2017A&ARv..25....2P}
P.~{Padovani}, D.~M. {Alexander}, R.~J. {Assef}, B.~{De Marco}, P.~{Giommi},
  R.~C. {Hickox} et~al., \emph{{Active galactic nuclei: what's in a name?}},
  \href{https://doi.org/10.1007/s00159-017-0102-9}{\emph{\aapr} {\bfseries 25}
  (2017) 2} [\href{https://arxiv.org/abs/1707.07134}{{\ttfamily 1707.07134}}].

\bibitem{2016ATel.8879....1P}
G.~{Piano}, A.~{Bulgarelli}, M.~{Tavani}, I.~{Donnarumma}, C.~{Pittori},
  F.~{Verrecchia} et~al., \emph{{AGILE detection of a gamma-ray flare from the
  FSRQ PKS 2023-07}}, {\emph{The Astronomer's Telegram} {\bfseries 8879} (2016)
  }.

\bibitem{2016ATel.8932....1C}
S.~{Ciprini} and {Fermi Large Area Telescope Collaboration}, \emph{{Fermi-LAT
  detection of a GeV gamma-ray flare from the blazar PKS 2023-07}}, {\emph{The
  Astronomer's Telegram} {\bfseries 8932} (2016) }.

\bibitem{2018A&C....22....9L}
J.-P. {Lenain}, \emph{{FLaapLUC: A pipeline for the generation of prompt alerts
  on transient Fermi-LAT {\ensuremath{\gamma}}-ray sources}},
  \href{https://doi.org/10.1016/j.ascom.2017.11.002}{\emph{Astronomy and
  Computing} {\bfseries 22} (2018) 9}
  [\href{https://arxiv.org/abs/1709.04065}{{\ttfamily 1709.04065}}].

\bibitem{2009APh....32..231D}
M.~{de Naurois} and L.~{Rolland}, \emph{{A high performance likelihood
  reconstruction of {\ensuremath{\gamma}}-rays for imaging atmospheric
  Cherenkov telescopes}},
  \href{https://doi.org/10.1016/j.astropartphys.2009.09.001}{\emph{Astroparticle
  Physics} {\bfseries 32} (2009) 231}
  [\href{https://arxiv.org/abs/0907.2610}{{\ttfamily 0907.2610}}].

\bibitem{2015arXiv150900794M}
T.~{Murach}, M.~{Gajdus} and R.~D. {Parsons}, \emph{{A Neural Network-Based
  Monoscopic Reconstruction Algorithm for H.E.S.S. II}}, {\emph{arXiv e-prints}
  (2015) arXiv:1509.00794} [\href{https://arxiv.org/abs/1509.00794}{{\ttfamily
  1509.00794}}].

\bibitem{2009ApJ...697.1071A}
W.~B. {Atwood}, A.~A. {Abdo}, M.~{Ackermann}, W.~{Althouse}, B.~{Anderson},
  M.~{Axelsson} et~al., \emph{{The Large Area Telescope on the Fermi Gamma-Ray
  Space Telescope Mission}},
  \href{https://doi.org/10.1088/0004-637X/697/2/1071}{\emph{\apj} {\bfseries
  697} (2009) 1071} [\href{https://arxiv.org/abs/0902.1089}{{\ttfamily
  0902.1089}}].

\bibitem{2015ApJS..218...23A}
F.~{Acero}, M.~{Ackermann}, M.~{Ajello}, A.~{Albert}, W.~B. {Atwood},
  M.~{Axelsson} et~al., \emph{{Fermi Large Area Telescope Third Source
  Catalog}}, \href{https://doi.org/10.1088/0067-0049/218/2/23}{\emph{\apjs}
  {\bfseries 218} (2015) 23}
  [\href{https://arxiv.org/abs/1501.02003}{{\ttfamily 1501.02003}}].

\bibitem{2019arXiv190210045T}
{The Fermi-LAT collaboration}, \emph{{Fermi Large Area Telescope Fourth Source
  Catalog}}, {\emph{arXiv e-prints} (2019) arXiv:1902.10045}
  [\href{https://arxiv.org/abs/1902.10045}{{\ttfamily 1902.10045}}].

\bibitem{2004ApJ...611.1005G}
N.~{Gehrels}, G.~{Chincarini}, P.~{Giommi}, K.~O. {Mason}, J.~A. {Nousek},
  A.~A. {Wells} et~al., \emph{{The Swift Gamma-Ray Burst Mission}},
  \href{https://doi.org/10.1086/422091}{\emph{\apj} {\bfseries 611} (2004)
  1005} [\href{https://arxiv.org/abs/astro-ph/0405233}{{\ttfamily
  astro-ph/0405233}}].

\bibitem{2005SSRv..120..165B}
D.~N. {Burrows}, J.~E. {Hill}, J.~A. {Nousek}, J.~A. {Kennea}, A.~{Wells},
  J.~P. {Osborne} et~al., \emph{{The Swift X-Ray Telescope}},
  \href{https://doi.org/10.1007/s11214-005-5097-2}{\emph{\ssr} {\bfseries 120}
  (2005) 165} [\href{https://arxiv.org/abs/astro-ph/0508071}{{\ttfamily
  astro-ph/0508071}}].

\bibitem{2005A&A...440..775K}
P.~M.~W. {Kalberla}, W.~B. {Burton}, D.~{Hartmann}, E.~M. {Arnal}, E.~{Bajaja},
  R.~{Morras} et~al., \emph{{The Leiden/Argentine/Bonn (LAB) Survey of Galactic
  HI. Final data release of the combined LDS and IAR surveys with improved
  stray-radiation corrections}},
  \href{https://doi.org/10.1051/0004-6361:20041864}{\emph{\aap} {\bfseries 440}
  (2005) 775} [\href{https://arxiv.org/abs/astro-ph/0504140}{{\ttfamily
  astro-ph/0504140}}].

\bibitem{2016arXiv161205560C}
K.~C. {Chambers}, E.~A. {Magnier}, N.~{Metcalfe}, H.~A. {Flewelling}, M.~E.
  {Huber}, C.~Z. {Waters} et~al., \emph{{The Pan-STARRS1 Surveys}},
  {\emph{arXiv e-prints} (2016) arXiv:1612.05560}
  [\href{https://arxiv.org/abs/1612.05560}{{\ttfamily 1612.05560}}].

\bibitem{2011MNRAS.410.2556D}
A.~{Dom{\'\i}nguez}, J.~R. {Primack}, D.~J. {Rosario}, F.~{Prada}, R.~C.
  {Gilmore}, S.~M. {Faber} et~al., \emph{{Extragalactic background light
  inferred from AEGIS galaxy-SED-type fractions}},
  \href{https://doi.org/10.1111/j.1365-2966.2010.17631.x}{\emph{\mnras}
  {\bfseries 410} (2011) 2556}
  [\href{https://arxiv.org/abs/1007.1459}{{\ttfamily 1007.1459}}].

\bibitem{2005NIMPA.551..493R}
W.~A. {Rolke}, A.~M. {L{\'o}pez} and J.~{Conrad}, \emph{{Limits and confidence
  intervals in the presence of nuisance parameters}},
  \href{https://doi.org/10.1016/j.nima.2005.05.068}{\emph{Nuclear Instruments
  and Methods in Physics Research A} {\bfseries 551} (2005) 493}
  [\href{https://arxiv.org/abs/physics/0403059}{{\ttfamily physics/0403059}}].

\bibitem{2018Sci...362.1031F}
{Fermi-LAT Collaboration}, S.~{Abdollahi}, M.~{Ackermann}, M.~{Ajello}, W.~B.
  {Atwood}, L.~{Baldini} et~al., \emph{{A gamma-ray determination of the
  Universe's star formation history}},
  \href{https://doi.org/10.1126/science.aat8123}{\emph{Science} {\bfseries 362}
  (2018) 1031} [\href{https://arxiv.org/abs/1812.01031}{{\ttfamily
  1812.01031}}].

\bibitem{2016ARA&A..54..725M}
G.~. {Madejski} and M.~{Sikora}, \emph{{Gamma-Ray Observations of Active
  Galactic Nuclei}},
  \href{https://doi.org/10.1146/annurev-astro-081913-040044}{\emph{\araa}
  {\bfseries 54} (2016) 725}.

\bibitem{2016ApJ...830...94F}
J.~D. {Finke}, \emph{{External Compton Scattering in Blazar Jets and the
  Location of the Gamma-Ray Emitting Region}},
  \href{https://doi.org/10.3847/0004-637X/830/2/94}{\emph{\apj} {\bfseries 830}
  (2016) 94} [\href{https://arxiv.org/abs/1607.03907}{{\ttfamily 1607.03907}}].

\bibitem{2019arXiv190202291M}
M.~{Meyer}, J.~D. {Scargle} and R.~D. {Blandford}, \emph{{Characterizing the
  Gamma-Ray Variability of the Brightest Flat Spectrum Radio Quasars Observed
  with the Fermi LAT}},
  \href{https://doi.org/10.3847/1538-4357/ab1651}{\emph{\apj} {\bfseries 877}
  (2019) 1} [\href{https://arxiv.org/abs/1902.02291}{{\ttfamily 1902.02291}}].

\end{thebibliography}\endgroup

\end{document}